\documentclass[10pt,aps,prl,twocolumn,english,superscriptaddress]{revtex4-2}
\usepackage{amssymb}
\usepackage{color}
\usepackage{graphicx}
\usepackage{bbold}
\usepackage{bm}
\usepackage{amsmath}
\usepackage{amsfonts}
\usepackage{amssymb}
\usepackage{braket}
\usepackage{upgreek}
\usepackage{xcolor}
\usepackage{fancyhdr}
\usepackage{float}
\usepackage{mathrsfs}
\usepackage{siunitx}
\usepackage{textcmds}
\usepackage{makecell}
\usepackage{orcidlink}
\usepackage{CJKutf8}
\usepackage{tikz}
\usepackage{txfonts}

\renewcommand{\selectlanguage}[1]{}

\usepackage[normalem]{ulem}

\definecolor{burgundy}{RGB}{97,0,35}
\definecolor{marine}{RGB}{4,46,96}
\definecolor{greendark}{RGB}{3,53,0}
\definecolor{bggreen}{RGB}{185,230,70}
\definecolor{greyed}{RGB}{170, 170, 170}
\definecolor{myblue}{RGB}{0, 40, 140}
\definecolor{blankcolor}{RGB}{220, 220, 230}

\usepackage{hyperref}
\hypersetup{%
    colorlinks=true,
    urlcolor=marine,
    linkcolor=marine,
    citecolor=marine
    }
\usepackage{lipsum}

\newcommand{\Sr}[1]{$^{#1}$Sr}

\newcommand{\RedMotTransitionShort}{${^1\mathrm{S}_0} - {^3\mathrm{P}_1}$}

\newcommand{\Pone}{$^3$P$_1$}
\newcommand{\Pzero}{$^3$P$_0$}

\newcommand{\Szero}{$^1$S$_0$}
\newcommand{\Sone}{$^3$S$_1$}

\DeclareSIUnit\gauss{G}

\definecolor{bggreen}{RGB}{185,230,70}
\definecolor{myblue}{RGB}{0, 40, 140}

\begin{document}

\renewcommand{\figurename}{Fig.}
\renewcommand{\tablename}{Table}

\title{Coherent Three-Photon Excitation of the Strontium Clock Transition}

\author{Junyu He (何君鈺)\,\orcidlink{0000-0003-2137-7248}}
\affiliation{Van der Waals-Zeeman Institute, Institute of Physics, University of Amsterdam, Science Park 904, 1098XH Amsterdam, The Netherlands}

\author{Benjamin Pasquiou\,\orcidlink{0000-0001-5374-2129}}
\affiliation{Van der Waals-Zeeman Institute, Institute of Physics, University of Amsterdam, Science Park 904, 1098XH Amsterdam, The Netherlands}
\affiliation{QuSoft, Science Park 123, 1098XG Amsterdam, The Netherlands}
\affiliation{CNRS, UMR 7538, LPL, F-93430, Villetaneuse, France}
\affiliation{Laboratoire de Physique des Lasers, Université Sorbonne Paris Nord, F-93430, Villetaneuse, France}

\author{Rodrigo González Escudero\,\orcidlink{0000-0001-7315-6733}}
\affiliation{Van der Waals-Zeeman Institute, Institute of Physics, University of Amsterdam, Science Park 904, 1098XH Amsterdam, The Netherlands}

\author{Sheng Zhou (\mbox{周晟})\,\orcidlink{0000-0002-8230-6577}}
\affiliation{Van der Waals-Zeeman Institute, Institute of Physics, University of Amsterdam, Science Park 904, 1098XH Amsterdam, The Netherlands}

\author{Mateusz Borkowski\,\orcidlink{0000-0003-0236-8100}}
\affiliation{Van der Waals-Zeeman Institute, Institute of Physics, University of Amsterdam, Science Park 904, 1098XH Amsterdam, The Netherlands}
\email{mateusz@cold-molecules.com}

\author{Florian Schreck\,\orcidlink{0000-0001-8225-8803}}
\email[]{threePhotonTransfer@strontiumBEC.com}
\affiliation{Van der Waals-Zeeman Institute, Institute of Physics, University of Amsterdam, Science Park 904, 1098XH Amsterdam, The Netherlands}
\affiliation{QuSoft, Science Park 123, 1098XG Amsterdam, The Netherlands}

\date{\today}

\begin{abstract}
We demonstrate coherent three-photon excitation of the strontium clock transition with a contrast of 51(12)\% using a Bose-Einstein condensate. We follow it up with a demonstration of three-photon STIRAP-like transfer, overcoming the typical limitations of this technique to odd-level numbers. We also measure the two-body loss coefficient of \Sr{84} clock-state atoms. Our work constitutes an essential step towards outcoupling a continuous atom laser beam and provides a fast excitation mechanism for quantum simulation using bosonic alkaline-earth-like atoms.
\end{abstract}

\begin{CJK*}{UTF8}{min}
\maketitle
\end{CJK*}

Two-valence-electron atoms like Sr and Yb offer significant advantages over alkali atoms for quantum sensing \cite{ludlow_optical_2015, yu_gravitational_2011, graham_new_2013}, simulation \cite{gorshkov_two-orbital_2010,gerbier_gauge_2010} and computing \cite{morgado_quantum_2021, daley_quantum_2008}. These advantages originate in a narrow-line transition enabling laser cooling to the microkelvin scale, a ground state free of electronic magnetic moment (\Szero), metastable excited states ($^3$P$_{0,2}$), single-photon Rydberg transitions from those states, and an ultranarrow optical clock transition (\Szero~-- \Pzero). The clock transition provides the frequency reference optical atomic clocks \cite{ludlow_optical_2015, aeppli_clock_2024}, enables elegant atom interferometric gravitational-wave detectors \cite{yu_gravitational_2011, graham_new_2013}, and qubits \cite{morgado_quantum_2021, pucher_fine-structure_2024, unnikrishnan_coherent_2024}. 

\begin{figure}[b]
	\centering
	\includegraphics[width=\columnwidth]{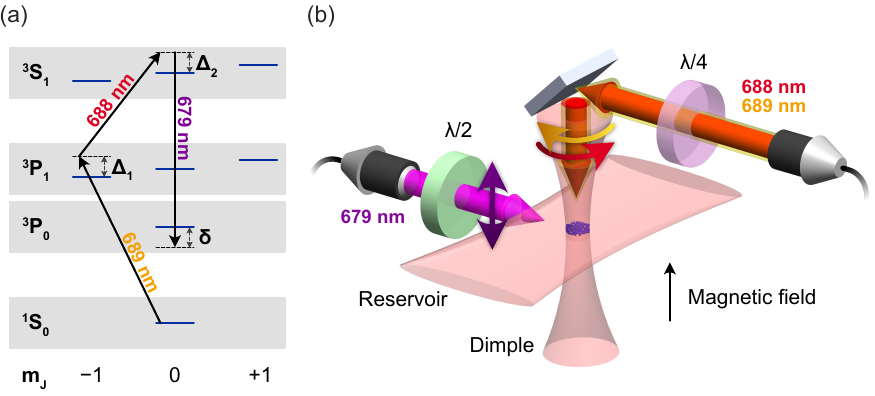}
	\caption{
 (a) Levels and transitions relevant for three-photon transfer. We add a non-zero magnetic field to lift the degeneracy between $m_J$ sub-levels. By controlling the polarizations of the beams, we drive the three-photon transfer through specific sub-levels that avoid destructive interferences between potential transfer paths. $\Delta_1$: one-photon detuning for 689-nm. $\Delta_2 - \Delta_1$: one-photon detuning for 688-nm. $\delta$: three-photon detuning. (b) Geometry of dipole traps and three-photon transfer beams. Colored arrows denote photon polarizations.}
	\label{Fig1}
\end{figure}

Bosonic strontium provides experimental opportunities not offered by the fermionic isotope. Only bosons allow the creation of Bose-Einstein condensates (BECs) and atom lasers, which are indispensable tools for quantum simulation~\cite{bloch_quantum_2012} and promise quantum sensing based on atom interferometry~\cite{robins_atom_2013,alonso_cold_2022}. The clock transition enable the construction of simplified clocks, atom interferometric gravitational-wave antennas, or the outcoupling of an atom laser beam from a steady-state BEC~\cite{chen_continuous_2022}. This transition can be opened by applying a magnetic field \cite{barber_direct_2006, taichenachev_magnetic_2006, akatsuka_optical_2008, origlia_towards_2018}, but this technique only yields small Rabi frequencies unless one uses impractically high magnetic fields or light intensities. Instead, we implement a three-photon Raman transfer scheme \cite{barker_three-photon_2016, hong_optical_2005} based on a system of four levels: the ground state \Szero, clock state \Pzero, and two intermediate states, \Pone~and \Sone~[Fig.~\ref{Fig1}(a)].

Here we demonstrate a coherent three-photon Raman transfer of Sr atoms on the clock transition. First, using a \Sr{84} BEC as a starting point, we observe Rabi oscillations between the ground state \Szero~and the excited clock state \Pzero. Second, we use the same scheme to demonstrate Stimulated Raman Adiabatic Passage (STIRAP), overcoming its usual limitation to odd level numbers~\cite{vitanov_stimulated_2017}. Third, in a thermal gas, we measure the two-body loss coefficient for collisions of clock-state atoms.

Throughout this work we either use a pure BEC or a thermal gas of $^{84}$Sr. Briefly, we load atoms from a red Sr magneto-optical trap (MOT) into a crossed optical dipole trap, extinguish the MOT, and perform evaporative cooling \cite{bennetts_steady-state_2017, chen_continuous_2022}. The dipole trap consists of a horizontally-propagating elliptical reservoir beam with waists $\SI{120}{\micro \meter}$ horizontally and $\SI{14.5}{\micro \meter}$ vertically, and a vertically-propagating dimple trap with a beam waist of $\SI{27}{\micro \meter}$ [Fig.~\ref{Fig1}(b)]. Both use 1070-nm light. We obtain pure BECs of about 10$^5$ atoms. When a $\mu$K-level thermal gas is needed, we stop evaporation early.

For our three-photon Raman scheme [Fig.~\ref{Fig1}(a)] it is critical that we maintain coherence between the three lasers driving it. We generate the 689-nm light for the $^1$S$_0$-$^3$P$_1$ transition using an injection-locked diode laser, which amplifies light from our MOT laser. The MOT laser has a full width at half-maximum of approximately $\SI{1.5}{\kilo \hertz}$ after locking to a piezo cavity. We stabilize the frequency of this laser to better than \SI{0.5}{\kilo \hertz} by locking the cavity length using saturated-absorption spectroscopy  \cite{madison_degenerate_2014}. We produce the 688-nm beam for the $^3$P$_1$-$^3$S$_1$ transition using a diode laser whose low output power of $\SI{2.5}{\milli \watt}$ is then increased to $\SI{13}{\milli \watt}$ using an injection-locked laser. We generate the 679-nm beam for the downward $^3$S$_1$-$^3$P$_0$ transition using a diode laser that delivers $\SI{1.2}{\milli \watt}$ of power. We lock the frequencies of the three light sources using a transfer cavity (finesse $>9000$), placed under vacuum on vibration-isolating Viton rods. We stabilize the length of the cavity to the reference 689-nm light using a piezo. We then lock the 688-nm and 679-nm lasers to the transfer cavity using the Pound-Drever-Hall technique. The three laser fields are locked to better than {$\SI{15}{\kilo \hertz}$} of each other.

\begin{figure}[tb]
	\centering
	\includegraphics[width=1.0\columnwidth]{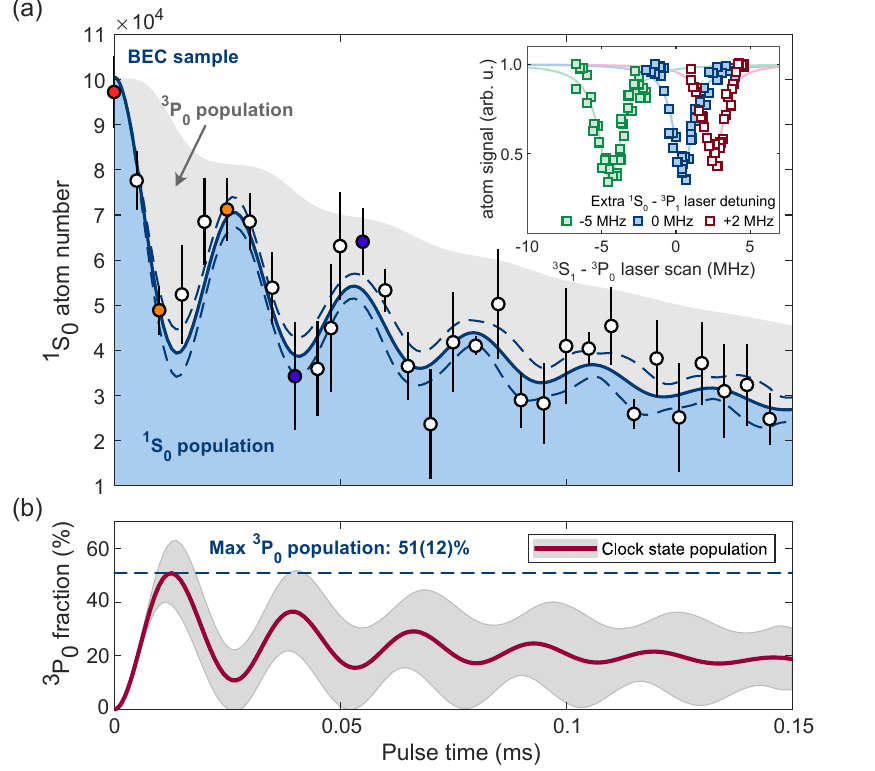}
	\caption{(a) Coherent three-photon transfer between \Szero~and \Pzero~state, starting from a BEC. The experimental data (open black circles) is fitted with a dynamics model using optical Bloch equations [Eq.~(\ref{eq:obes})]. The error bars represent a 68\% confidence interval calculated using the t-distribution~\cite{Taylor1994}. We determine an effective Rabi frequency of $2 \pi\,\times\,$29.8(1.7) kHz. Dashed lines denote the error of the fit. The light gray and blue shades describe the atomic population in the \Pzero~state and \Szero~state, respectively. Inset: the three-photon line observed by scanning the frequency of the third, \Sone-\Pzero, photon. As we change the frequency of the first, \Szero-\Pone~photon, we see a corresponding shift in the resonance position. (b) Clock state population according to the fitted optical Bloch equations model. At a pulse length of 12.5~$\mu$s we reach a maximum population of 51(12)\%. The dark gray shade represents the uncertainty of the fit. }
	\label{Fig2}
\end{figure}

The magnetic sublevels of the two intermediate states complicate our four-level picture and could lead to destructive interference between all the possible excitation pathways. To overcome this, we add a 1.78-G, vertically aligned magnetic field as a quantization axis, and carefully select the polarizations and propagation directions of the three laser beams [Fig.~\ref{Fig1}(a)]. We combine the 689-nm and 688-nm beams with orthogonal polarizations in the same polarization-maintaining fiber and propagate them parallel to the magnetic field, focused to a waist of $\SI{23}{\micro \meter}$ at the atom position. Using a quarter-wave plate we switch to circular polarizations to respectively drive the $\sigma^-$ and $\sigma^+$ transitions to $^3$P$_1$ $m_J = -1$ and then $^3$S$_1$ $m_J = 0$. The 679-nm beam is vertically-polarized, has a waist of $\SI{47}{\micro \meter}$, and propagates perpendicular to the magnetic field, driving the $\pi$ transition to the $^3$P$_0$ $m_J = 0$ state.

Our choice of laser detunings and intensities is a balancing act between achieving a large effective Rabi frequency, reducing off-resonant scattering, and accommodating the dual function of these lasers in the experimental scheme. The 689-nm laser operates with an on-resonant Rabi frequency of $\Omega_1 = 2 \pi\,\times\,\SI{0.87}{\mega \hertz}$ and at a detuning of $\Delta_1 = 2 \pi \times \SI{40}{\mega \hertz}$, which is significantly larger than the natural linewidth ($2 \pi\,\times\,7.5$\,kHz~\cite{Borkowski2014}) of this transition. The 688-nm laser also serves as a transparency beam~\cite{Stellmer2013LaserCoolingToBEC, chen_continuous_2022, Sonderhouse2020ThermodynDeepFermiGas}, which requires both a large detuning ($\Delta_2 - \Delta_1 = 2 \pi \times \SI{7340}{\mega \hertz}$) and large intensity ($\Omega_2 = 2 \pi \times \SI{1583}{\mega \hertz}$). The 679-nm laser completes the Raman scheme with a detuning of $2 \pi\,\times\,\SI{7380}{\mega \hertz}$ and $\Omega_3 = 2 \pi \times \SI{31.1}{\mega \hertz}$. We work within these constraints so that in the future we could employ this scheme to outcouple a continuous atom laser beam from a steady-state BEC~\cite{chen_continuous_2022}.

We first demonstrate a coherent three-photon Rabi oscillation [Figure~\ref{Fig2}(a)]. After preparing the BEC we apply the three-photon pulse by first turning on the 688-nm laser beam and a few $\SI{}{\micro \second}$ later the 689-nm and 679-nm beams. We measure the remaining \Szero~atom number by absorption imaging after $\SI{24}{\milli \second}$ of free-flight expansion. By varying the pulse duration we clearly observe several Rabi oscillations within the first $150\,\mu$s. After the first oscillation (orange) we see a revival of 23(7)\% of the initial population (red), and after the second (blue), a revival of 31(11)\%.

To determine the clock-state atom fraction, we fit an effective optical Bloch equations model~\cite{metcalf1999laser} (Appendix A). We choose a model that treats the two-body loss coefficient $\beta_{ee}$ as a fit parameter to avoid making assumptions about the clock-state cloud shape. We determine a Rabi frequency $\Omega = 2 \pi\,\times\,${29.8(1.7)\,kHz} and an effective detuning $\delta = 2 \pi\,\times\,${22.9(2.3)\,kHz}. The atom population oscillates between the ground and clock states with an effective frequency $\sqrt{\Omega^2+\delta^2} = 2 \pi\,\times\,$37.6(1.9)\,kHz. We find a strong correlation (correlation coefficient $0.98$) between $\beta_{ee} = 14(19)\times10^{-11}$~cm$^{3}$\,s$^{-1}$ and the decoherence rate $\Gamma_{\rm coh} = 2 \pi\,\times\,${5.6(6.5)}\,kHz which precludes their joint determination from this dataset; we instead determine $\beta_{ee}$ below using a thermal gas. With this fitted model, we further evaluate the transfer efficiency as a function of pulse length [Fig.2(b)]. We achieve a maximum transfer of 51(12)\% from the ground state to the clock state at a pulse length of 12.5~$\mu$s. Our other fitted models corroborate this determination.

To better understand the decoherence processes described by $\delta$ and $\Gamma_{\rm coh}$, we employ a Monte-Carlo calculation (Appendix B). Briefly, we find three main sources of decoherence: the inhomogeneity of the effective Rabi frequency [Eq.~(\ref{eq:omega_eff})] due to the Raman beam waists, light shift broadening due to trapping and Raman lasers, and the finite Raman laser linewidths. The effective Rabi frequencies vary as much as 20\% throughout the sample. This is responsible for the rapid loss of coherence. The light shifts add about 4.8\,kHz of variability to the laser detunings seen by the atoms. Lastly, the finite combined laser linewidth limits the clock-state transfer efficiency. Our Monte-Carlo calculations align best with the experimental data for a combined linewidth between $10$\,kHz and $30$\,kHz, in agreement with our laser characterization. The efficiency cannot be improved by simply changing the laser detuning.

\begin{figure}[tb]
	\centering
	\includegraphics[width=1.0\columnwidth]{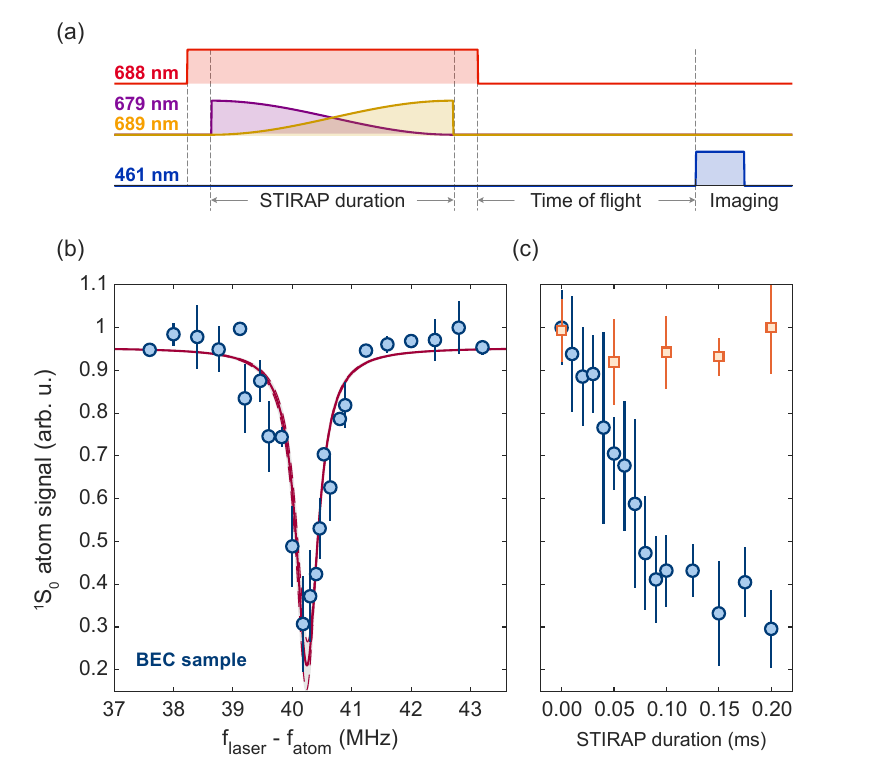}
	\caption{STIRAP of \Szero~atoms to the \Pzero~clock state, starting from a BEC. (a) We execute STIRAP pulses by varying the intensities of the 679-nm and 689-nm lasers, but keeping the 688-nm power constant. We then detect the remaining $^1$S$_0$ atom number by time-of-flight imaging. (b) Remaining \Szero~atom number depending on 689-nm laser detuning for a STIRAP duration of {$\SI{150}{\micro \second}$}. Here $f_{\rm laser}$ is the frequency of the 689-nm laser and $f_{\rm atoms}$ is the frequency of the \RedMotTransitionShort~transition at zero magnetic field. (c) Circles: Remaining ground-state atoms as a function of STIRAP pulse duration. Atom loss saturates at about 150\,$\mu$s. Squares: same sequence, but with the 679-nm laser off, showing that all three lasers are needed to perform the transfer. }
	\label{Fig3}
\end{figure}

Next, we demonstrate a STIRAP-like transfer of \Szero~atoms to \Pzero~[Fig.~\ref{Fig3}(a)], which provides higher reliability than $\pi$ pulses. Conventionally STIRAP is perceived to be limited to odd level numbers~\cite{vitanov_stimulated_2017}. Here, we reduce the effective number of states from four to three by coupling the $^3$P$_1$ and $^3$S$_1$ states using the 688-nm laser operating at constant power. The STIRAP transfer starts by applying 679-nm light but not 689-nm light and then slowly inverting the power ratio between the 679-nm and 689-nm light. A dark eigenstate thus adiabatically evolves from the \Szero~state into the \Pzero~state. At the end of the transfer, all light is turned off, with the 688-nm one last. With a STIRAP pulse duration of $\SI{150}{\micro \second}$, the upper limit on the transfer efficiency we measured is 74(4)\%, based on a Lorentzian fit to \Szero~loss data [Fig.~\ref{Fig3}(b)]. The actual transfer efficiency will be lower, as clock-state atoms are lost through two-body collisions. Increasing the pulse duration further does not improve efficiency [Fig.~\ref{Fig3}(c)], likely due to the collisional losses and finite linewidths of our lasers.

We employ several sanity checks to ensure that we target the 5s5p \Pzero~state. \emph{i})~We test the resonance condition $
    f_{\rm 689} + f_{\rm 688} - f_{679} = f_{^1{\rm{S}}_0-^3{\rm{P}}_0}$.
We vary the frequency of the first photon by several MHz while taking spectra using the third photon frequency [Fig.~\ref{Fig2}(a), inset]. This results in a corresponding shift of the three-photon line, as expected. \emph{ii})~We measure the absolute frequencies of the three lasers using a wavemeter. We obtain the three-photon transition frequency $f_{^1{\rm{S}}_0-^3{\rm{P}}_0} = 429\,227\,720(3)$~MHz, in perfect agreement with the literature $f_{^1{\rm{S}}_0-^3{\rm{P}}_0} = 429\,227\,716.762(6)$~MHz~\cite{morzynski_absolute_2015,miyake_isotope_2019}. \emph{iii})~We verify that we indeed populate the $^3$P$_0$ state. After a $\pi$-pulse, we first blow away all \Szero~atoms. We then recover a signal of \Szero~atoms by using the 679-nm and 688-nm lasers to two-photon transfer the \Pzero~atoms to the \Pone~state, which then decays to \Szero. \emph{iv})~We verify that the coherent behavior shown in Fig.~\ref{Fig2}(a) stems from a three-photon, rather than a two-photon transition. In principle the Rabi oscillations could stem from coupling to the bright $^3$S$_1$ state and the loss from subsequent decay to the metastable $^3$P$_{0,2}$ states. To demonstrate that also the 679-nm laser is crucial, we repeat our STIRAP experiment without turning on that laser. We see no atom loss~[squares in Fig.~\ref{Fig3}(c)]. \emph{v})~We compare the fitted Rabi frequency $\Omega = 2 \pi\,\times\,$ 29.8(1.7) kHz to the predictions of a four-level dressed-state model, where
\begin{equation}
    \Omega = \left|\frac{\Omega_{1} \Omega_{2} \Omega_{3}}{4 \Delta_{1} (\Delta_{2} - \Delta_{1}) - \Omega_{2}^2} \right|= 2\pi \times 32.2 (5.0)\,{\rm kHz},
    \label{eq:omega_eff}
\end{equation} assuming a 10\% power calibration uncertainty. We find perfect consistency.

\begin{figure}[b]
	\centering
	\includegraphics[width=1.0\columnwidth]{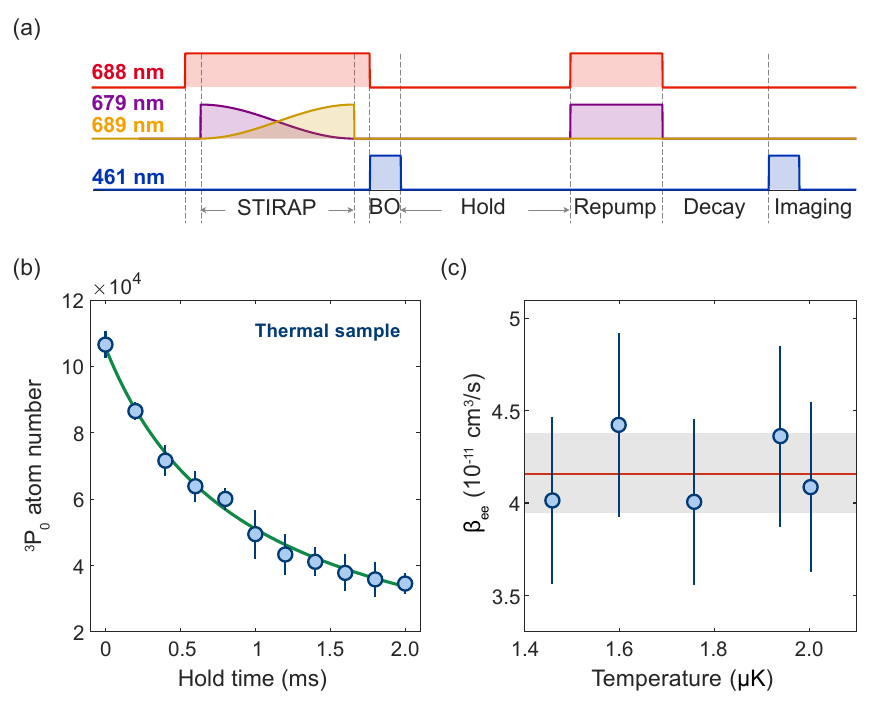}
	\caption{Measurements of two-body inelastic losses for \Pzero~atoms in thermal gases with densities ranging from 10$^{12}$ to 10$^{13}$ cm$^{-3}$. (a) We STIRAP-transfer the atoms to the clock state, purge the remaining atoms with a blow-out (BO) pulse and then hold the atoms in the dark (except for the trapping light) for collisions to occur. To detect the remaining atoms, we repump them to the $^3$P$_1$ state, let them decay to the ground state and then image. (b) Example \Pzero~atom number variation with hold time. The solid green line shows the fit with Eq.~(\ref{eq:2bodyloss}). The sample temperature is $\SI{1.76}{\micro \kelvin}$, and the fitted two-body loss coefficient is $\beta_{ee} = \SI{4.0(5)e-11 }{\cubic \centi \meter  \per \second}$. (c)~Two-body collisional loss coefficients $\beta_{ee}$ for different temperatures. The average across all datasets gives the two-body loss coefficient $\beta_{ee} = \SI{4.2(2)e-11 }{\cubic \centi \meter \per \second}$ and is shown as a line.}
	\label{Fig4}
\end{figure}

Last, we turn our attention to two-body inelastic collisions between clock-state atoms. Clock-state collisions are universally lossy and were previously studied in calcium, strontium and ytterbium \cite{bishof_inelastic_2011, bouganne_clock_2017,franchi_state-dependent_2017, halder_inelastic_2013, lisdat_collisional_2009,ludlow_cold-collision-shift_2011, schafer_spin_2017, tomita_dissipative_2019, traverso_inelastic_2009}. While clock-state atoms are metastable, the atomic interactions as the atoms approach allow for radiative decay to the many lower-lying molecular states~\cite{jones_ultracold_2006}. These are undetectable in our experiment. The collisional loss rates are typically on the order of $10^{-11}$\,cm$^3$/s and prevent the production of clock-state BECs via evaporative cooling. 

We measure the two-body loss coefficient of \Pzero~\Sr{84} atoms and its dependence on the sample temperature (Fig.~\ref{Fig4}). To isolate two-body loss from three-body loss we use a dilute thermal sample rather than a BEC. We again start with a STIRAP pulse [Fig.~\ref{Fig4}(a)], but immediately follow it up with a {$\SI{15}{\micro \second}$}-long blow-out pulse to remove the remaining ground-state atoms from the trap. Then, we hold the atoms in the dark (except for the trapping light) to allow for two-body collisions to occur. To detect the remaining clock-state atom fraction, we simply reuse two of our Raman lasers. We tune the 679-nm $^3$P$_1$-$^3$S$_1$ laser by 40 MHz to achieve a Raman resonance condition between the $^3$P$_0$ and $^3$P$_1$ states. From there, the atoms spontaneously decay to the ground state. The entire process has a 37(4)\% efficiency, likely limited by differential light shifts~\footnote{In principle, this could be also done with an extra pair of on-resonant 679-nm and 707-nm lasers~\cite{ludlow_optical_2015}.}. Fig.~\ref{Fig4}(b) shows the remaining atom number as the hold time is varied; we observe loss on a timescale of $\sim \SI{2}{\milli \second}$. 

To determine the two-body loss coefficient $\beta_{ee}$, we model the time variation of the atom number $N(t)$ via
\begin{equation}
\frac{d N(t)}{dt} = -\beta_{\mathrm{ee}} \frac{n_0}{N_0} N^2 (t).
\label{eq:2bodyloss}
\end{equation}
The initial density $n_0 = N_0 (\bar{\omega}^2 m/4 \pi k_b T)^{3/2}$ assumes a Maxwell-Boltzmann distribution for a temperature $T$. The mean trap frequency $\Bar{\omega}=(\omega_x \omega_y \omega_z)^{1/3}$; we measured the individual frequencies $\omega_{x, y, z}$ by observing dipole oscillations of a ground-state thermal gas. 
We assume that the clock-state atoms inherit the shape of the initial ground-state cloud because the hold time is too short for the atoms to thermalize and adapt to the new trapping potential. We independently measure the temperature of the \Szero~cloud before the transfer to the \Pzero~state via time-of-flight expansion. The initial densities of the thermal gas samples are prepared to be on the order of 10$^{12}$--10$^{13}$ cm$^{-3}$.
Even though our model only includes two-body losses, it fully describes the experimental data in Fig.~\ref{Fig4}(b). Other loss mechanisms can be ignored: trap-induced one-body loss is about $\SI{0.036}{\per \second}$; three-body losses should also occur at much longer time scales. Our data is insensitive to the latter; we attempted to add them to the model but that failed to improve the quality of the fit.

\begin{figure}
	\centering
	\includegraphics[width=\columnwidth]{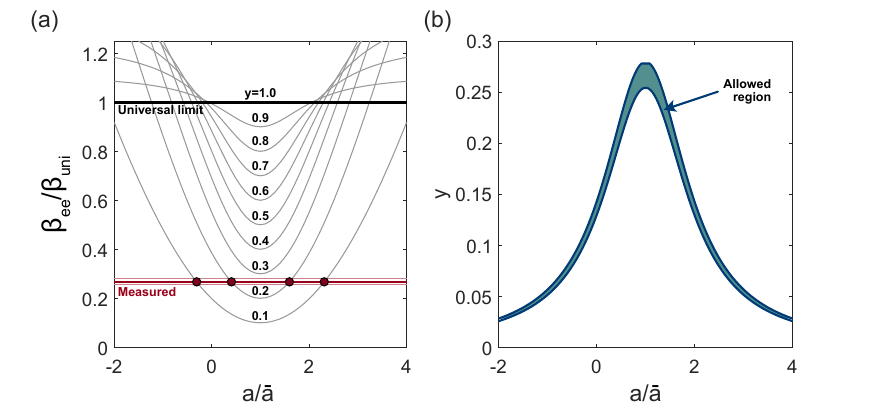}
	\caption{(a) In the universal model~\cite{Idziaszek2010}, two-body loss coefficients depend on the scattering length~$a$ and a short-range loss probability~$y$. Our measured loss coefficient $\beta_{\rm ee} = 4.2(2)\times 10^{-11}$~cm$^3$/s is four times smaller than the universal limit $\beta_{\rm uni}$. (b) Set of reduced scattering lengths $a$ and short range loss probabilities $y$ consistent with the measured loss coefficient shown in panel (a). } 
    \label{Fig5}
\end{figure}

We measure the two-body loss coefficients for temperatures ranging from $1.4$ to $\SI{2.0}{\micro \kelvin}$ [Fig.~\ref{Fig4}(c)]. The error bars are dominated by contributions from trap frequency stability, $^3$P$_0$ atom number calibration and to a smaller extent temperature. Within uncertainty, the two-body loss rate is constant in this temperature range, and we obtain an average loss coefficient $\beta_{ee} = \SI{4.2(2)e-11 }{\cubic \centi \meter \per \second}$. We compare our result to previous measurements using other isotopes in Table~\ref{tbl:beta_ee}. A value of similar magnitude to ours has been reported for a bulk thermal gas of \Sr{88} \cite{traverso_inelastic_2009}. Measurements for \Sr{87} and \Sr{88} confined to a 1D lattice result in smaller values \cite{bishof_inelastic_2011, lisdat_collisional_2009}, likely because of dimensional effects~\cite{lisdat_collisional_2009, petrov_interatomic_2001}.

\begin{table}[b]
\begin{center}
\begin{ruledtabular}
\caption{Variation of $s$-wave $^3$P$_0$-$^3$P$_0$ loss coefficients across strontium isotopes. \label{tbl:beta_ee}}
\begin{tabular}{ c r r r } 
Isotope & $\beta_{ee}$ (cm$^3$~s$^{-1}$) & Ref. & Note \\
\hline
\Sr{84} & $4.2(2)\times 10^{-11}$ & This work & Bulk \\
\Sr{88} & $2.3(1.4)\times 10^{-11}$ & \cite{traverso_inelastic_2009} & Bulk\\ 
\Sr{87} & $0.7(4)\times 10^{-11}$ & \cite{bishof_inelastic_2011} & 1D optical lattice\\ 
\Sr{88} & $4.0(2.5)\times 10^{-12}$ & \cite{lisdat_collisional_2009} & 1D optical lattice \\ 
\end{tabular}
\end{ruledtabular}
\end{center}
\end{table}

Our measured inelastic loss rates can be understood within the universal model~\cite{Idziaszek2010}. Due to wavefunction symmetry, clock state collisions in $^{84}$Sr can only occur in even partial waves. For our microkelvin sample this is effectively limited to the $s$-wave, where the universal model predicts the loss rate coefficient to be independent of temperature. Indeed, our data in Fig.~\ref{Fig4}(c) shows no thermal variation. We calculate the universal limit, $\beta_{\rm uni} = 4 h \bar a /\mu = 15.7(2)\times 10^{-11}$~cm$^3$/s using the mean scattering length $\bar{a}= 2^{-\frac{3}{2}}\frac{\Gamma(3/4)}{\Gamma(5/4)}(2\mu C_6/ \hbar^2)^\frac{1}{4} = 81.8$ a.u., the van der Waals coefficient $C_6 = 5360(200)$~a.u.~\cite{zhang_spectroscopic_2014}, and the reduced mass $\mu$ equal to half the atomic mass of strontium. This is four times larger than our $\beta_{\rm ee} = 4.2(2)\times 10^{-11}$~cm$^3$/s. The universal model describes the two-body loss coefficient via the (here unknown) scattering length $a$ and the short-range loss probability $y$ [Fig.~\ref{Fig5}(a)]. Here, having only one measured quantity, we cannot independently determine both $a$ and $y$. For any scattering length $a$, there is a matching short-range loss probability $y$ that matches the measured $\beta_{\rm ee}$ [Fig.~\ref{Fig5}(b)]. However, we notice that $y$ reaches a maximum when the scattering length $a$ equals $\bar a$, which allows us to determine an upper limit on $y$. We conclude that clock-state $^{84}$Sr collisions are not universal with the short-range loss probability $y < 0.268(13)$.

In conclusion, we have demonstrated a coherent transfer of bosonic Sr atoms between ground and excited clock states, starting with a BEC. We have reached a Rabi frequency of 29.8(1.7)\,kHz, a transfer efficiency of 51(12)\% and a coherence time that is several times the Rabi oscillation period. Further, we have shown our three-photon scheme allows a STIRAP-like transfer of up to 74(4)\% efficiency, as determined from a loss feature. Finally, we measured the two-body loss coefficient for $s$-wave collisions of clock-state~$^{84}$Sr atoms.

Compared to directly driving the clock transition at low magnetic fields, the three-photon transfer enables much higher Rabi frequencies, and could allow probing the clock transition in non-magic traps. It should be possible to improve the transfer efficiency by reducing the laser linewidths, ensuring a homogeneous illumination, and using a magic trap. Already as it is, our scheme could be used to outcouple a cw atom laser beam from a continuously replenished BEC~\cite{chen_continuous_2022}. So far, such a BEC only exists in Sr, where the outcoupling cannot be done by conventional means \cite{robins_atom_2013}: Sr is nonmagnetic, so radiofrequency transitions cannot be used; directly driving the clock transition with a high magnetic field is impractical, as it would interfere with the low magnetic field necessary for the MOT that continuously replenishes the BEC; it is also held in a deep dipole trap, so spilling is not an option either. One could use our scheme to coherently transfer atoms to a state for which the dipole trap can easily be engineered to be non-trapping, resulting in a free-falling atom laser beam. Upon excitation, the atoms receive a small net momentum; this can be used to regulate the momentum transfer from null~\cite{ryabtsev_doppler-_2011, barker_three-photon_2016} up to the three photon momenta combined. A large momentum kick helps outcouple a well-collimated atom laser beam \cite{robins_achieving_2006}, while suppressing the momentum transfer would help create a metastable-state BEC useful for quantum simulation~\cite{barker_three-photon_2016}. In quantum computing, our technique could push clock qubit gate times to under a microsecond~\cite{morgado_quantum_2021}.

\begin{acknowledgments}

\textit{Acknowledgments---}We are grateful to Z. Idziaszek, R. Ciuryło, P. Thekkeppatt, R. Spreeuw and Z. Guo for helpful discussions. We thank the RbSr team for lending the laser head and N. Grenier for building the 679-nm laser source. This work was supported by the Dutch National Growth Fund (NGF), as part of the Quantum Delta NL programme.

\textit{Author contributions---}J.H. led the experimental effort, with support from R.G.E. and S.Z., performed the data analysis and wrote the manuscript. M.B. contributed the theoretical analysis and cowrote the paper. B.P. conceived the use of the 3-photon Raman scheme as an atom laser outcoupler. J.H., B.P. and F.S. guided the effort. J.H., B.P., M.B. and F.S. discussed the results and F.S. acquired the funding.

\textit{Note---} During the completion of this work, we became
aware of concurrent work on the three-photon transition in \Sr{88}~\cite{carman2024collinear}.

\end{acknowledgments}

\onecolumngrid

\begin{center}
{{\bf{}End Matter}}
\end{center}

\twocolumngrid

\textit{Appendix A: Least-squares analysis of the Rabi oscillation data ---}
To determine the $^3$P$_0$ population during the three-photon transfer, we employ a set of optical Bloch equations where we treat our Raman process as an effective two-level system~\cite{metcalf1999laser}:
\begin{eqnarray}
    \frac{d}{dt} \rho_{gg} & = 
        & \frac{1}{2} i \Omega \left (\rho_{eg} - \rho_{ge} \right), \nonumber \\
    \frac{d}{dt} \rho_{ee} & = 
        & \frac{1}{2} i \Omega \left (\rho_{ge} - \rho_{eg} \right) 
        -\left(\Gamma_e + g \beta_{ee} \bar n \rho_{ee} \right) \rho_{ee},  \nonumber \\
    \frac{d}{dt} \rho_{ge} & = 
        & \frac{1}{2} i \Omega \left (\rho_{ee} - \rho_{gg} \right) 
        -\frac{1}{2}\left( \Gamma_e + \Gamma_{\rm coh} + g \beta_{ee} \bar n \rho_{ee} + 2i \delta \right) \rho_{ge}, \nonumber \\
    \frac{d}{dt} \rho_{eg} & = 
        & \frac{1}{2} i \Omega \left (\rho_{gg} - \rho_{ee} \right) 
        -\frac{1}{2}\left( \Gamma_e +\Gamma_{\rm coh} + g \beta_{ee} \bar n \rho_{ee} - 2i \delta \right) \rho_{eg}. \nonumber \\
        \label{eq:obes}
\end{eqnarray}
The density matrix $\rho$ describes the atom populations and coherences. The subscripts $g$ and $e$ denote the ground state ($^1$S$_0$) and clock state ($^3$P$_0$). The fitted parameter $\Omega$ is the three-photon Rabi frequency. The decoherence rate $\Gamma_{\rm coh}$ and an effective detuning $\delta$ account for decoherence effects. The fixed parameter $\Gamma_e = 2\pi\,\times\,$54\,Hz accounts for the slow off-resonant scattering from the 679-nm Raman beam~\footnote{Other off-resonant scattering sources are negligible at these timescales.}. We account for two-body losses of clock-state atoms via the $\beta_{ee}$ term, with the symmetry factor $g=1$ for a clock-state BEC and $g=2$ for a thermal gas. The initial average density of ground-state atoms, $\bar n(t=0) = 2.78\times10^{14}\,$cm$^{-3}$, assumes a Thomas-Fermi profile.

\begin{table}[b]
\caption{Least-squares fitting parameters for the optical Bloch equations [Eq.~(\ref{eq:obes})] modelling Rabi oscillation data in Fig.~\ref{Fig2}. The three fits differ in their treatment of two-body losses of clock-state atoms. \label{tbl:lsq}}
\begin{ruledtabular}
    \begin{tabular}{l c r r r}
        Parameter           & Unit                          & Fit A     & Fit B     & Fit C         \\
        \hline
        $g$                 &                               & 1         & 2         & 2             \\
        dof                 &                               & 27        & 27        & 26            \\
        $\chi^2/{\rm{}dof}$ &                               & 1.75      & 1.24      & 0.91          \\
        $N_{\rm BEC}$       & $10^3$                        & 88.5(8.0) & 95.9(6.9) & 100.5(5.2)    \\
        $\Omega$            & $2\pi\,\times$~kHz            & 30.1(2.9) & 30.8(1.9) & 29.8(1.7)     \\
        $\delta$            & $2\pi\,\times$~kHz            & 22.4(3.8) & 21.3(2.8) & 22.9(2.3)     \\
        $\sqrt{\Omega^2 + \delta^2}$  & $2\pi\,\times$~kHz  & 37.5(3.2) & 37.4(2.2) & 37.6(1.9)     \\
        $\Gamma_{\rm coh}$  & $2\pi\,\times$~kHz            & 12.3(4.2) & 11.2(2.8) & 5.6(6.5)      \\
        $\beta_{\rm ee}$    & $10^{-11}$~cm$^{3}$s$^{-1}$   & fixed     & fixed     & 14(19)        \\
        max $\rho_{ee}$     & \%                            & 57.6(8.5) & 59.1(5.7) & 51(12)        \\
    \end{tabular}
\end{ruledtabular}
\end{table}

We perform three fits to the Rabi oscillation data shown in Fig.~\ref{Fig2} and list the fitted parameters in Table~\ref{tbl:lsq}. In the first fit (``A"), we assume that the density profile of clock-state atoms is the same as the initial Thomas-Fermi distribution for the ground-state-atom BEC, and that the clock-state cloud is itself a BEC ($g=1$). We also assume that the two-body loss factor $\beta_{ee}$ is equal to the one measured in a dilute thermal gas [Fig.~\ref{Fig4}(a)] and that all three-body losses can be ignored. This fit yields a $\chi^2/{\rm dof} = 1.75$ indicating a poor fit quality. In Fit B we forego the assumption of a clock-state BEC in favor of a thermal gas ($g=2$). This improves the fit quality to $\chi^2/{\rm dof} = 1.24$. In the final Fit C, we give up any assumptions on the two-body losses by making $\beta_{ee}$ a fitted parameter, further reducing $\chi^2/{\rm dof}$ to 0.91. We cautiously pick Fit C as our final result as it makes the fewest assumptions about cloud shape. Under Snedecor's F-test with a 5\% significance level, Fit C is indeed significantly better than Fit A, at $p=0.041$. However, the differences between Fit C and Fit B (and between Fit B and Fit A) are too small to be statistically significant. 

\begin{figure}[b]
    \includegraphics[width=1.0\linewidth]{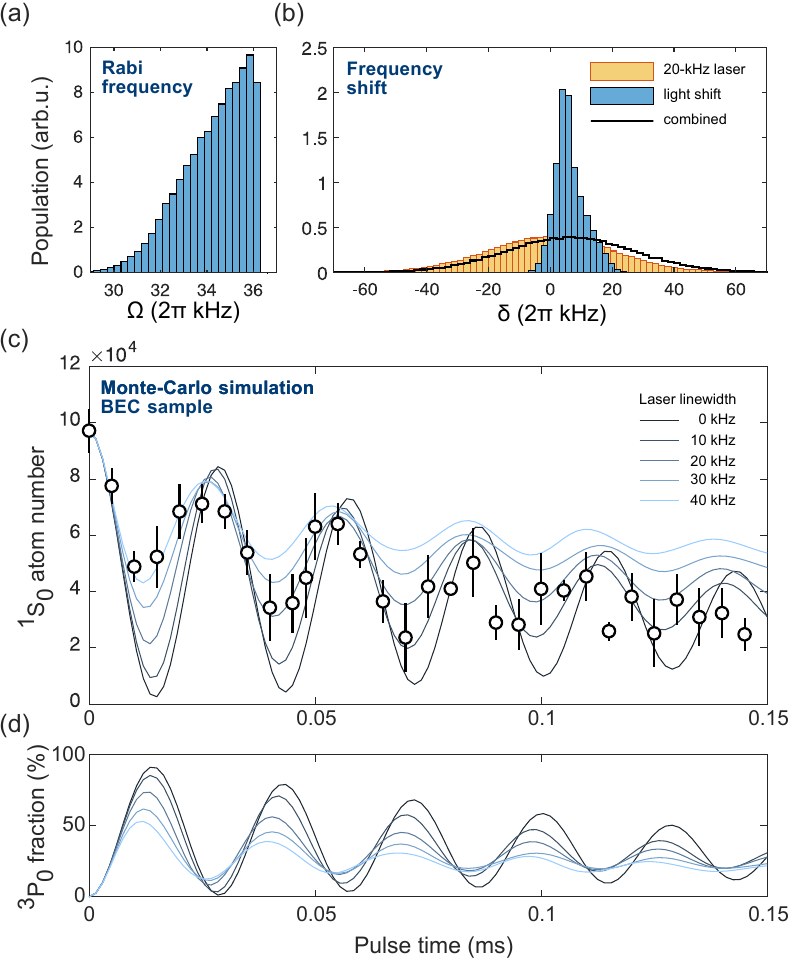}
    \caption{Monte-Carlo simulations of the population dynamics in our three-photon transfer. (a) Distribution of effective Rabi frequencies due to finite cloud and probe beam sizes. (b) Frequency shifts due to light shifts from 1070-nm trapping and 688-nm coupling lasers, and due to finite combined probe laser linewidth. (c) Monte-Carlo ground-state population dynamics compared to experimental data in Fig.~\ref{Fig2}. (d) Fractional clock-state population.}
    \label{Fig6}
\end{figure}

All determined parameters agree with each other to within one sigma across the three fits. Independent of the chosen model, the three-photon Rabi frequency $\Omega$ is consistently close to $2\pi\times30\,$kHz. The effective detuning, which partially accounts for the finite laser linewidth, hovers around $2\pi\times22\,$kHz. The Fit C value $\delta = 2\pi\times22.9(2.3)\,$kHz corresponds to a combined Gaussian linewidth of the three Raman lasers of $\sigma = \sqrt{\pi/2} \delta = 2\pi\times28.7(2.9)\,$kHz \footnote{One can understand the effective detuning as the mean absolute detuning of a laser that has a Gaussian distribution of frequency components. For a standard deviation $\sigma$ and zero mean, the mean absolute value is $\sqrt{2/\pi}\sigma$.}. Fits A and B give similar decoherence rates $\Gamma_{\rm coh}$ of about $2\pi\times12\,$kHz; in Fit C, the two-body loss coefficient $\beta_{ee}$ correlates strongly with the decoherence rate, making joint determination impossible. While Fit C does appear to favor a larger $\beta_{ee}$ than measured in a thermal gas, that could be explained by our omission of all three-body loss mechanisms, or two-body optical collisions. Lastly, all fits point to a maximum clock-state population $\rho_{ee}$ of over 50\%; our selected model yields a population of 51(12)\%.

\textit{Appendix B: Monte-Carlo simulations ---}
Our measured Rabi oscillation exhibits decoherence on a sub-millisecond timescale. To better understand its sources, we employ a Monte-Carlo simulation. As an initial condition, we consider a pure BEC of $^{84}$Sr atoms held by an approximately harmonic optical dipole trap ($\omega=2\pi\times \{165, 245, 168\}$\,Hz) formed at the intersection of the dimple and reservoir beams~[Fig.~1(b)]. We model the atomic density using a Thomas-Fermi profile~\cite{Pethick_Smith_2008}. For an initial population of $9.7\times10^4$ atoms and an $s$-wave scattering length of $122.7\,a_0$, the Thomas-Fermi radii $R_{\rm TF}=\{6.0,4.0,5.9\}~\mu$m and the peak density reaches $n_{\rm peak}=4.1 \times 10^{14}$\,cm$^{-3}$. 

The finite size of the atomic cloud coupled with the small waist of the Raman beams contributes to decoherence in a twofold manner. On the one hand, the variation in Raman laser intensities creates an inhomogeneity in the effective Rabi frequency. On the other hand, the light shift from the trapping lasers and the 688-nm laser creates an inhomogenous frequency shift.

The three Raman lasers are tightly focused at the atomic cloud. The 689-nm and 688-nm laser beams share the same fiber output, and are focused to a 23-$\mu$m $1/e^2$ beam waist $w$. For the 679-nm laser, $w=47\,\mu$m. Fig.~\ref{Fig6}(a) shows the variation in effective Rabi frequency as calculated for a sampling of atoms from the Thomas-Fermi distribution using Eq.~(\ref{eq:omega_eff}). We find that effective three-photon Rabi frequencies vary from over $2\pi\times{}36\,$kHz in the middle of the cloud to less than $2\pi\times30\,$kHz at its edges. Overall the distribution is strongly asymmetrical, with a mean of $2\pi\times34.3\,$kHz and standard deviation of $2\pi\times1.4\,$kHz.

The trapping light at 1070-nm and the $^3$P$_1$-$^3$S$_1$ transition laser at 688-nm both shift the effective $^1$S$_0$-$^3$P$_0$ resonance frequency. The 1070-nm dimple beam has a waist of 27\,$\mu$m, the tight axis of the sheet beam has a waist of 14.5\,$\mu$m. The differential polarizability at 1070-nm is $-65$\,a.u. While the beam power of the 688-nm laser is smaller than the trapping light, the large differential polarizability of 1442 a.u. causes this beam to make an appreciable contribution to the light shift. Overall we find that the distribution of effective light shifts [Fig.~\ref{Fig6}(b)] has a standard deviation of 4.8~kHz. Finally, we also include the finite laser linewidths as an effective combined Gaussian width $\sigma$. For $\sigma = 2\pi\times 20$~kHz the laser linewidth easily dominates the light shift.

Figure~\ref{Fig6}(c) shows the result of a Monte-Carlo simulation of ground-state population dynamics for several combined laser linewidths. Even assuming the probe lasers are perfectly on-resonance, we find that the inhomogeneity in Rabi frequencies and the frequency spread caused by the finite laser linewidth and light shift leads to a rapid loss of coherence consistent with our measurements. Qualitatively, we find that the simulations for laser linewidths between 10 kHz and 30 kHz match the experimental data best. This is roughly in agreement with the 23-kHz effective detunings determined from our fitted model. The clock state population [Fig. \ref{Fig6}(d)] also qualitatively reproduces that of our fitted model. We conclude that in the future we could improve our excitation efficiency by narrowing down the laser linewidths, implementing a magic-frequency trap to counteract the light shift, and by increasing the probe laser beam waists to homogenize the effective Rabi frequency. 

\bibliography{Sr}

\end{document}